# Electric field Control of Exchange Bias by Resistive Switching

L. J. Wei[1], Z. Z. Hu[2], Y. J. Wang[2], G. X. Du[2, *], Y. Yuan[1], J. Wang[1], H. Q. Tu[1], B. You[1, 5], S. M. Zhou[3], Y. Hu[4], J. Du[1, 5,*]

[1]*National Laboratory of Solid State Microstructures and Department of Physics, Nanjing University, Nanjing 210093, P. R. China*

[2]*Department of Mathematics and Physics, Nanjing university of posts and telecommunications, Nanjing 211167, P. R. China*

[3]*Department of Physics, Tongji University, Shanghai 200092, P. R. China*

[4]*College of Sciences, Northeastern University, Shenyang 110819, China*

[5]*Collaborative Innovation Center of Advanced Microstructures, Nanjing 210093, P. R. China*

We demonstrated an electric field controlled exchange bias (EB) effect accompanied with unipolar resistive switching behavior in the $Si/SiO_2/Pt/Co/NiO/Pt$ device. By applying certain voltages, the device displays obvious EB in high-resistance-state while negligible EB in low-resistance-state. Conductive filaments forming and rupture in the NiO layer but near the Co-NiO interface are considered to play dominant roles in determining the combined resistive switching and EB phenomena. This work paves a new way for designing multifunctional and nonvolatile magnetic-electrical random access memory devices.

---

*Authors to whom correspondence should be addressed. Electronic addresses: duguanxiang@njupt.edu.cn and jdu@nju.edu.cn

Exchange coupling between an antiferromagnet (AFM) and a ferromagnet (FM) can give rise to a unidirectional anisotropy known as the exchange bias (EB) effect. Although it has been widely used in spintronic applications [1], such as read heads or magnetic sensors, the inherent mechanism has not been completely understood. Recently, electric-field (E-field) control of magnetism has received growing interest due to the attractive merits of reduced power consumption, increased integration density and enhanced functionality in spintronic devices [2-4]. Among them, E-field control of EB in AFM/FM heterostructures could lead to deterministic 180° magnetic switching, which is of great significance for information storage [3]. Besides the potential applications, E-field control of EB could also offer a technique for exploring the EB mechanism.

To our knowledge, many earlier studies of E-field control of EB were performed on magnetoelectric/magnetic heterostructures involving single phase multiferroics materials, such as $Cr_2O_3$ [5-7], $YMnO_3$ [8] and $BiFeO_3$ [3,9,10], which are both ferroelectric/dielectric and antiferromagnetic below their Néel temperature. For examples, with $Cr_2O_3$ coupled to Pd-Co multilayers, He *et al* [6] claimed that it was possible to reversibly switch between two different EB polarity states at room temperature through the application of electric field and an additional applied magnetic field after magneto-electric field cooling. Wu *et al* [10] reported the creation of a multiferroic field effect device with a $BiFeO_3$ (antiferromagnetic-ferroelectric) gate dielectric and a $La_{0.7}Sr_{0.3}MnO_3$ (ferromagnetic) conducting channel that exhibits direct, bipolar electrical control of EB at quite low temperature (~ 5.5 K). However, in these studies, reversible electrical magnetization reversal has not been achieved because the modulation of EB field is less than the coercive field [6,8-10].

Very recently, the combination of AFM/FM exchange-biased systems and ferroelectric materials [11,12] gives a new pathway to achieve reversible E-field-controlled magnetization reversal by combining strain-mediated magnetoelectric coupling and EB. Liu et al [11] realized a magnetization switching by nearly 180° at room temperature in FeMn/Ni$_{80}$Fe$_{20}$/FeGaB/PMN-PT (lead zinc niobate-lead titanate) exchange-biased multiferroic system through E-field-tuned EB, however, this magnetization switching is irreversible. Chen et al [12] carefully investigated the angular dependence of E-field-controlled EB and magnetization reversal at room temperature in the exchange-biased Co$_{40}$Fe$_{40}$B$_{20}$ (CoFeB)/Ir$_{25}$Mn$_{75}$ (IrMn) heterostructures deposited on PMN-PT ferroelectric single crystal. They demonstrated that reversible electrical modulation of EB and magnetization reversal were obtained, which depend strongly on the angle between external magnetic field and the pinning direction. However, special angle and huge constant applied voltage (~ 400 V) are needed.

In this work, we report an alternative approach in achieving room-temperature reversible E-field control of EB in Si/SiO$_2$/Pt/Co/NiO/Pt heterostructures by virtue of resistive switching (RS) effect in the NiO layer. RS phenomenon refers to that reversible switching between a high-resistance state (HRS) and a low-resistance state (LRS) can be controlled by applying a certain electric field, which has been widely observed in various oxides when they are sandwiched by two metal electrodes [13-17]. Resistive random access memory (RRAM) based on the RS effect has attracted a lot of interest in recent years because of its remarkable characteristics, such as low power consumption, high operation speed, nondestructive readout, etc [18]. By applying appropriate voltages in the present Si/SiO$_2$/Pt/Co/NiO/Pt devices, larger on/off ratio (~10$^5$) could be obtained at room temperature with distinguished HRS and LRS, which correspond to distinct EB and negligible EB, respectively. The E-field control of EB in combination with RS in AFM/FM heterostructures paves a new way for designing magneto-electric random access memories (MERAMs). It is also promising for future development of multifunctional and non-volatile memory devices.

**Results**

In order to achieve both significant EB and good RS performance in the present devices, two NiO layers were fabricated with different degrees of oxidation, as mentioned above. For verifying this, the XPS depth profile characterization was performed on the sample. The XPS spectra were corrected using the maximum of the adventitious C 1s signal at 284.8 eV. It is noted that the XPS experiments were carried out on the sample surface region where the top Pt electrodes have not been covered, i.e. the separation areas between the neighboring Pt electrodes. The XPS spectra were recorded respectively after etching the sample for 300 s and 500 s with the etching rate evaluated to be about 0.1 nm/s, and the probing depth is between 5 nm and 10 nm. Therefore, the oxidation states of Ni in different NiO layers can be detected. As shown in Fig. 1(b), the top and bottom XPS panels are corresponding to NiO(2) and NiO(1) layers, respectively. After careful fitting, all the peaks can be distinguished. One of them for NiO(1) layer shows a metallic nickel ($Ni^0$-2p3/2) peak at the binding energy of 852.4 eV. The other two peaks for both NiO(1) and NiO(2) layer are located at the binding energies of 854.0 eV and 855.5 eV, which are corresponding to $Ni^{2+}$-2p3/2. Moreover, an obvious peak at 861.4 eV is recognized as the satellite peak for $Ni^{2+}$-2p$^{3/2}$. The XPS spectra analysis also indicate that the Ni ions with their valences larger than 2 are absent, eliminating the other nickel-oxides such as $Ni_2O_3$, $NiO_2$ in the sample. The most striking result is that $Ni^0$ atoms can be only observed in the NiO(1) layer while they are absent in the NiO(2) layer, indicating that the NiO(2) layer is almost fully oxidized while the NiO(1) layer still contains metallic Ni atoms besides the NiO compounds.

The bright-field cross-sectional TEM images of the Pt/Co/NiO multilayer film are shown in Fig. 2, which exhibit good stacking structure and sharp interfaces. As shown in Fig. 2 (a), there is no clear interface between NiO(1) and NiO(2) layers, which is possibly due to that they have the similar crystalline structure. The thicknesses of the Co and the NiO layers are about 6 and 68 nm, respectively, in good agreement with the nominal ones. Moreover, the NiO layer has a columnar structure with the width of the columnar grain of about 10 nm (see Supplementary Information S1). Fig. 2 (b) shows the selected area diffraction (SAD) pattern of the sample, which verifies the polycrystalline structure for the NiO layers. Furthermore, from the high resolution TEM (HRTEM) images shown in Fig. 4(c)-4(f) with respect to the labeled locations in the NiO layers, the interplanar spacings are measured to be about 0.2400/0.2389 and 0.2080/0.2079 nm, corresponding to the NiO (111) and (200) planes, respectively. These results are in good consistent with those obtained by the XRD patterns (see Supplementary Information S2).

Figure 3 shows a typical unipolar RS behavior in the Si/SiO$_2$/Pt/Co/NiO/Pt device. That is, the Pt/Co/NiO/Pt device can be switched from a HRS to a LRS without changing the voltage polarity. The current-voltage (*I-V*) loops of the memory cells were studied by dc voltage sweep measurements, and the results are displayed in both linear (Fig. 3(a)) and semi-logarithmic scales (Fig. 3(b)). The device was first underwent a "forming" process by increasing the voltage from 0 V to about 4.3 V (see the inset of Fig. 3(a)) and then in LRS. A current compliance (Icc) of 1 mA was applied to avoid permanent dielectric breakdown of the device. After the 'forming' process, the applied voltage was reduced to zero. Afterwards, with increasing the voltage gradually from zero, the LRS changed to the HRS when the voltage was above the 'RESET' value of about 1.4 V with the resistance increased abruptly from ~ 40 Ω to ~ $10^8$ Ω. And then the voltage was reduced to zero again. With increasing the applied voltage from zero again, the device changed back to the LRS when the voltage was above the 'SET' value of about 3.5 V without changing the voltage polarity. We also investigated the reliability and stability of the two-state RS characteristic of the device. Fig. 3(c) shows the cycle number dependence of the device's resistance at LRS and HRS, respectively. Although the 200 cycles shown in the endurance test in Fig. 3(c) are encouraging, practical devices will require thousands or millions of switching cycles. As shown in Fig. 3(d), the resistances at both LRS and HRS keep almost unchanged under a fixed applied voltage of 0.1 V even after $10^4$ seconds, which demonstrates a good retention time performance for the device. Moreover, the on/off ratio ($R_{HRS}/R_{LRS}$) is larger than $10^5$ (see Fig. 3(c) and Fig. 3(d)), allowing one to distinguish the two resistance states easily in a real working device. The mechanism responsible for the above RS behavior will be addressed in the latter parts. To further understand the resistance switching behavior, several conduction models have been used to fit the *I-V* data (see Supplementary Information S3). The fitting results demonstrate that the Ohmic conduction and space charge limited conduction (SCLC) mechanisms are responsible for the LRS and HRS, respectively.

Figure 4 shows the typical *M-H* loops for different cells on the sample measured by focus MOKE. Before the *M-H* loop measurements, the cells were set at different resistance states by applying certain voltages and all the top Pt electrodes were removed by IBE. In Fig. 4(a), the red circles represent the laser spot probing areas and the letters of 'O', 'L' and 'H' indicate that the probing cell is in original state (O-S, without applying any voltages), LRS state and HRS, respectively. Moreover, the red circle with a letter of 'Q' in the middle denotes the probing area overlapping LRS and O-S. As shown in Fig. 4(b), clear left shift of the *M-H* loop along the magnetic field axis can be observed, confirming the establishment of EB in the original state (O-S). The quantity of the EB field is defined as $H_E = -(H_{CL}+H_{CR})/2$, where $H_{CL}$ and $H_{CR}$ are the coercive fields for the descent and ascent branches of the *M-H* loop, respectively. Then, $H_E$ is determined to be 20 Oe for O-S. The inset of Fig. 4(b) exhibits the *M-H* loop of a 5 nm Co film without neighboring NiO layers, which shows that $H_E$ is zero and $H_C$ is about 16 Oe. As shown in Fig. 4(c), the *M-H* loops indicate that $H_E$ is about 19.1 Oe for HRS, which is almost the same as that for O-S, and nearly zero for LRS. Moreover, $H_C$ is calculated to be about 23 Oe for LRS, which is obviously larger than that (16 Oe) of a single Co layer with the same thickness, suggesting that weak interfacial coupling still exists between the Co and NiO layers for LRS. Comparing the *M-H* loop for HRS with that for LRS, one can find that the descending branch moves rightwards significantly whereas the ascending branch keeps almost unchanged. Therefore, when the laser spot is located at the 'Q' region which spans over LRS and O-S, the resultant *M-H* loop will be superposition of the HRS- and LRS-loops, which is verified in Fig. 4(d). As shown in the inset of Fig. 4(d), because the laser spot covers about 50% LRS and 50% O-S region, the HRS- and LRS-loops seem to be comparable. However, when the laser spot is deviated from this position, the superposed *M-H* loop will change significantly with different proportions of the two loops (see Supplementary Information S4). Moreover, it needs to be emphasized that we have measured a lot of cells preset at different resistance states (HRS or LRS) and the above similar EB results could be obtained (see Supplementary Information S5).

**Discussion**

The correlation between the RS behavior and EB effect can be understood based on the conductive filamentary path model [13,19], according to which the conductive filament (CF) is produced during the 'Forming' or 'SET' step under high electric fields, and ruptured during the 'RESET' step due to high current densities. It is well accepted that NiO is a p-type wide bandgap semiconductor, where the major injected carrier is a hole [20,21]. The Joule heating effect, which facilitates the redox reaction, is most serious at local locations around the anodic interface where hole injection occurs due to diffusive transport [22]. Oxygen loss must occur at highly localized areas where holes are injected. However, when oxygen ions are lost by the redox reaction at the anodic interface, Ni interstitials ($Ni_i^{..}$) can be formed through the following reaction:

$$\text{NiO} \rightarrow Ni_i^{..} + 2e^- + \frac{1}{2}O_2 \qquad (1)$$

The generated $Ni_i^{..}$ ions may drift/diffuse toward the cathode interface. Under this circumstance, the point of holes injection might be a source for supplying $Ni_i^{..}$ to the rest of the NiO layer. When $Ni_i^{..}$ generation becomes severe enough, they meet together and form tiny metallic Ni filaments. Finally, the percolated metallic Ni chains develop the CFs in the form of dendrite-like or random network [22]. After such a 'Forming' or 'SET' step, the device changes from O-S to LRS.

As shown in the top panel of Fig. 5(a), it is noted that the NiO layer in O-S contains some nanometer-sized conductive domains at highly localized areas, such as the grain boundaries (dash lines) [19]. Furthermore, first-principle calculation suggest low migration energy at the grain boundary [23]. Therefore, it is believed that Ni atoms moving along the grain boundaries provide the necessary atoms to form the CFs across the electrodes under the voltage, as displayed in the top panel of Fig. 5(b) or Fig. 5(c). So, the CFs will be generally formed at the columnar grain boundaries (blue vertical lines) [19,24]. However, when the CFs grow towards the cathode, they may pass through the original grain boundaries to form new grain boundaries (blue non-vertical lines). Similar results can be also found in previous reports [19,24,25]. When the device is in O-S, due to interfacial exchange coupling between Co and NiO, EB will be established easily. Although some Ni atoms or clusters reside in the bottom NiO layer, which has been verified by the XPS results shown in Fig. 1(b), they can possibly depress the device's insulation level but hardly damage the AFM structure of the interfacial NiO layer and thus have negligible influence on EB. However, when the device is positively biased and transformed from O-S to LRS, the Ni CFs will be formed throughout the entire NiO layer and a portion of them reach the NiO-Co interface, which will most possibly damage the AFM structure there, leading to reduced or even disappeared EB. The corresponding magnetic moment arrangements of the Ni and Co atoms around the NiO-Co interface in different resistance states are illustrated in the bottom panels in Figs. 5(a)-5(c).

After a 'RESET' step, the device will enter HRS because the CFs has been ruptured. As mentioned above, the filament grows from the anode interface to the cathode interface. Therefore, the weakest part of the filament is most possibly formed near the cathode, where the localized rupture and recovery of the filament occur [26]. This is because the Ni filament near the cathode has the highest resistance so that the Joule heating effect is most serious there. The most likely way of rupturing the percolated conducting channel is to move some of the $O^{2-}$ ions from the NiO region nearby the Ni filament to the Ni filament portion near the cathode [26]. In other words, the Ni filament near the Co-NiO interface will be oxidized to NiO again during the rupture process. Therefore, as shown in Fig. 5(c), the AFM structure near the Co-NiO interface has been almost recovered after the 'RESET' step, leading to appearance of EB again as that in O-S. Since the RS behavior, i.e. the switching between 'SET' and 'RESET' processes, is reversible, the E-Field control of EB is also reversible.

In summary, Si/SiO$_2$/Pt/Co/NiO/Pt devices have been fabricated by magnetron sputtering, in which both unipolar RS behavior and EB effect can be well controlled by electric field at room temperature. By applying certain voltages, the device can be intentionally set at 'HRS' with obvious exchange bias and at 'LRS' with negligible EB. Considering the conductive filaments forming and rupture in the NiO layer and near the Co-NiO interface as well, all the correlated RS and EB phenomena can be well explained. This work provides a new approach to achieve reversible E-field controlled EB, which could pave a new way to realize multifunctional and nonvolatile MERAM devices with extremely low energy consumption.

**Methods**

**Device fabrication:** The Pt/Co/NiO/Pt multilayer films were deposited by magnetron sputtering at room temperature on Si (100) substrate with native oxide on the surface. Before the film deposition, the commercial Si wafers were diced into about 1 cm × 1 cm pieces as substrates and cleaned by proper procedures. The sputtering targets of Pt, Co and NiO were purchased commercially and their purities are all larger than 99.99%. A schematic illustration of the sample's stacking structure is shown in Fig. 1(a). The base pressure was lower than $5.0 \times 10^{-6}$ Pa and the Ar pressure was kept at 0.3 Pa during the deposition for all the metal films. A Pt film of about 50 nm was first deposited on the substrate as the bottom electrode, followed by a 6 nm Co layer made by dc sputtering. Afterwards, the first NiO layer ($t_1 \sim 20$ nm) was deposited by rf sputtering with pure Argon gas. In order to increase the insulation level of the film, the second NiO layer ($t_2 \sim 50$ nm) was made by reactive rf sputtering with mixed gases of Argon and Oxygen ($P_{Ar} : P_{O_2} = 5 : 2$). For simplicity, the first and second NiO layers are denoted as NiO(1) and NiO(2), respectively. During deposition of the NiO layers, the sputtering power and the working gas pressure were maintained at 80 W and 0.5 Pa, respectively. Finally, for making the top electrode, another Pt layer of about 50 nm was deposited on the top NiO layer using shadow mask and the diameter of the top electrode is about 250 μm. Note that all the thicknesses mentioned above are nominal. To establish EB of the Co/NiO bilayer, a constant magnetic field of about 200 Oe was applied parallel to the film plane during the film deposition and no further field cooling was carried out.

**Characterizations:** The RS behavior of the sample was characterized by a Keithley-2400 meter with the wires' connection illustrated in the top panel of Fig. 1(a). Note that the positive bias is applied with the current flowing from the top Pt electrode to the NiO/Co layer and then to the bottom Pt electrode and the negative bias is applied with the current flowing in the opposite direction. The in-plane magnetic hysteresis (*M-H*) loops were measured by a commercial focused magneto-optic Kerr effect (MOKE, NanoMOKE 3) magnetometer with the magnetic field applied in the film plane and parallel to the incident plane of light as well, as shown in the bottom panel of Fig. 1(a). The MOKE system utilized a 660 nm red laser light with a spot size of about 200 μm. In order to let the probe light reach the Co layer to fulfill the MOKE measurement, the top Pt electrodes were removed by ion beam etching (IBE) prior to the measurements. X-ray photoemission spectroscopy (XPS, Thermo-scientific K-Alpha) measurements were carried out in an ultrahigh vacuum system using Al Kα radiation as the X-ray source. The crystal structure of sample was characterized by X-ray diffraction (XRD, Bruker D8AA25X) using Cu $K_\alpha$ radiation ($\lambda = 0.154$ nm). The microstructure was also characterized by a JEM-2100 transmission electron microscope (TEM) with 200 kV accelerating voltage. All the measurements and characterizations were performed at room temperature.

**Acknowledgements**

This work was supported by National Basic Research Program of China (2014CB921101), National Key Research and Development Program of China (2016YFA0300803), National Natural Science Foundations of China (Nos. 51471085, 51331004, 11174131, 61274102).

**Author contributions**

J.D. and S.Z. initiated the study. L.W. prepared the samples by magnetron sputtering. L.W. performed the *I-V* measurements analyzed the results. Z.H., L.W. and G.D. performed the FMOKE measurements and analyzed the results. L.W. performed the XRD, XPS and TEM characterizations and analyzed the results with J.D. L.W. performed the magnetic simulation and analyzed the results. L.W. and J.D. prepared the manuscript. All the authors contributed to discussion of the project and revision of the manuscript.

**Additional information**

The authors declare no competing financial interests. The details of the characterizations and calculations were shown in the supplementary information. Correspondence should be addressed to J. D. and G. X. D.

**Competing financial interests**

The authors declare no competing financial interests.

**Figure Legends**

**Fig. 1 Electric, magneto-optical measurement geometry and XPS spectra.** **(a)** Schematic illustration of the stacking structure and *I-V* measurement of the Si/SiO$_2$/Pt/Co/NiO(1)/NiO(2)/Pt sample (top panel) and MOKE measurement geometry after removal of the top Pt electrodes (bottom panel). **(b)** XPS spectra for NiO(1) (bottom panel) and NiO(2) (top panel).

**Fig. 2 Microstructure characterization by TEM.** The cross-sectional TEM image **(a)** and SAD pattern **(b)** of the Pt/Co/NiO multilayer film. **(c)**-**(f)** The HRTEM images corresponding to the locations denoted by 'c, d, e, f' in **(a)**.

**Fig. 3 Resistive switching performance characterization.** Typical *I-V* curves of the Pt/Co/NiO/Pt devices in both linear scale **(a)** and semi-logarithmic scale **(b)**. The inset of **(a)** is an enlarged view for the red dashed box. The arrows in **(a)** and **(b)** indicate the voltage sweeping directions. The cycle number **(c)** and retention time **(d)** dependences of the device's resistance at HRS and LRS under an applied voltage of 0.1 V.

**Fig. 4 Electric field control of exchange bias.** **(a)** Schematic illustration for different laser spot detection locations on the sample. **(b)**, **(c)** and **(d)** show the *M-H* loops where the device is in O-S, HRS/LRS, and Q, respectively. The inset of **(b)** displays the *M-H* loop of a Pt(50 nm)/Co(6 nm)/Pt(50 nm) film.

**Fig. 5 The mechanism responsible for the correlation between resistive switching and exchange bias.** Top panels schematically illustrate the 'Forming', 'SET' and 'RESET' processes when the positively biased Pt/Co/NiO/Pt device is in O-S (a), LRS (b) and HRS (c), respectively. Bottom panels exhibit the corresponding magnetic moment arrangements of the Ni and Co atoms around the NiO-Co interface when the device is in O-S (a), LRS (b) and HRS (c), respectively.

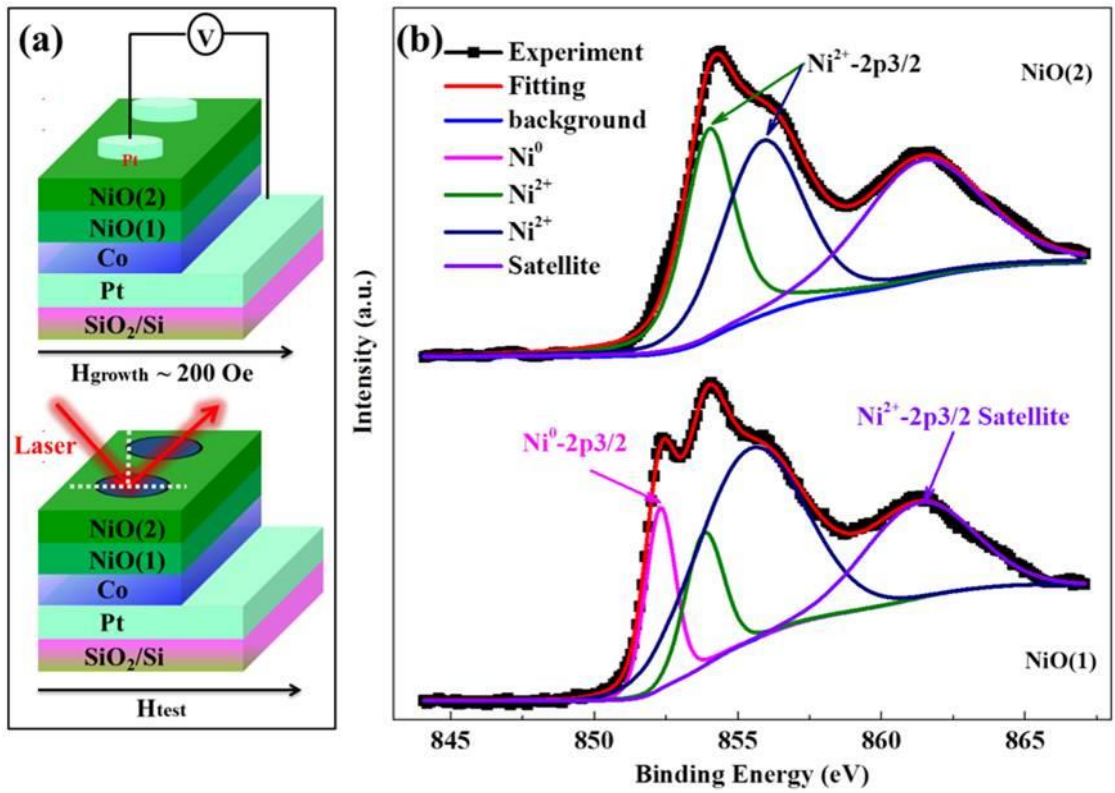

Fig. 1 Wei LJ et al

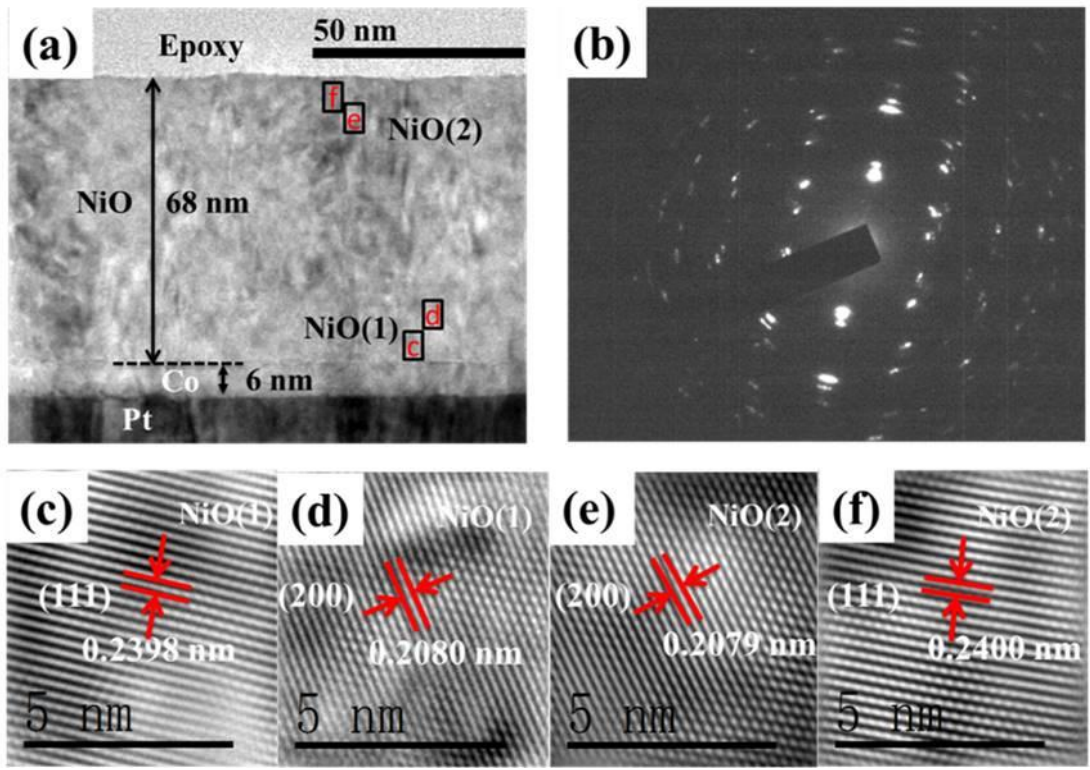

**Fig. 2 Wei LJ et al**

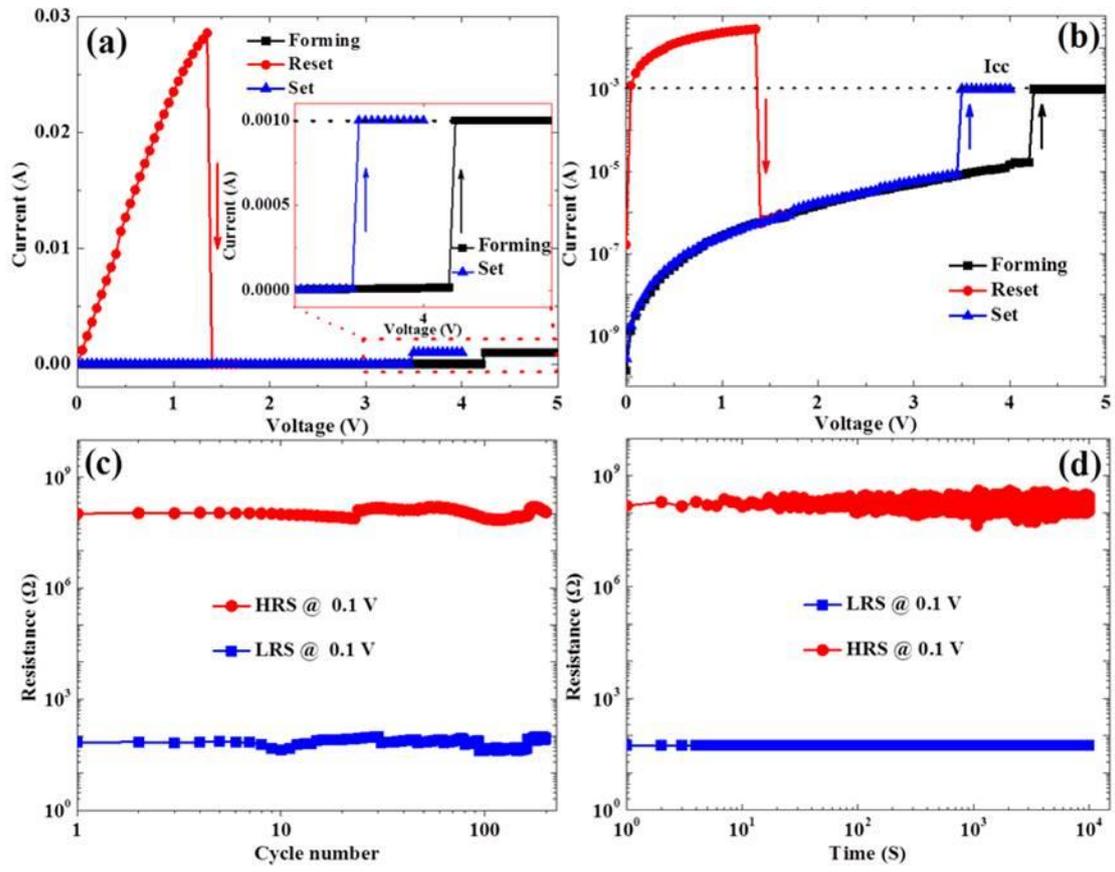

**Fig. 3 Wei LJ et al**

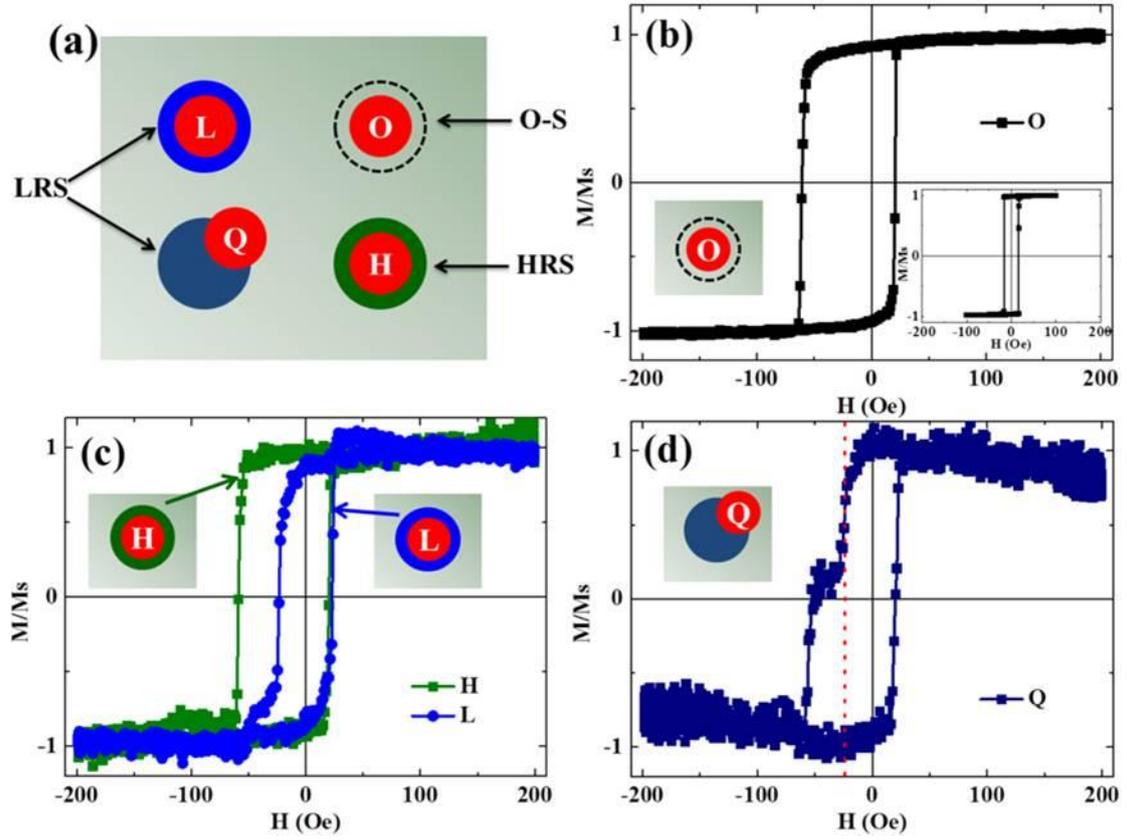

**Fig. 4 Wei LJ et al**

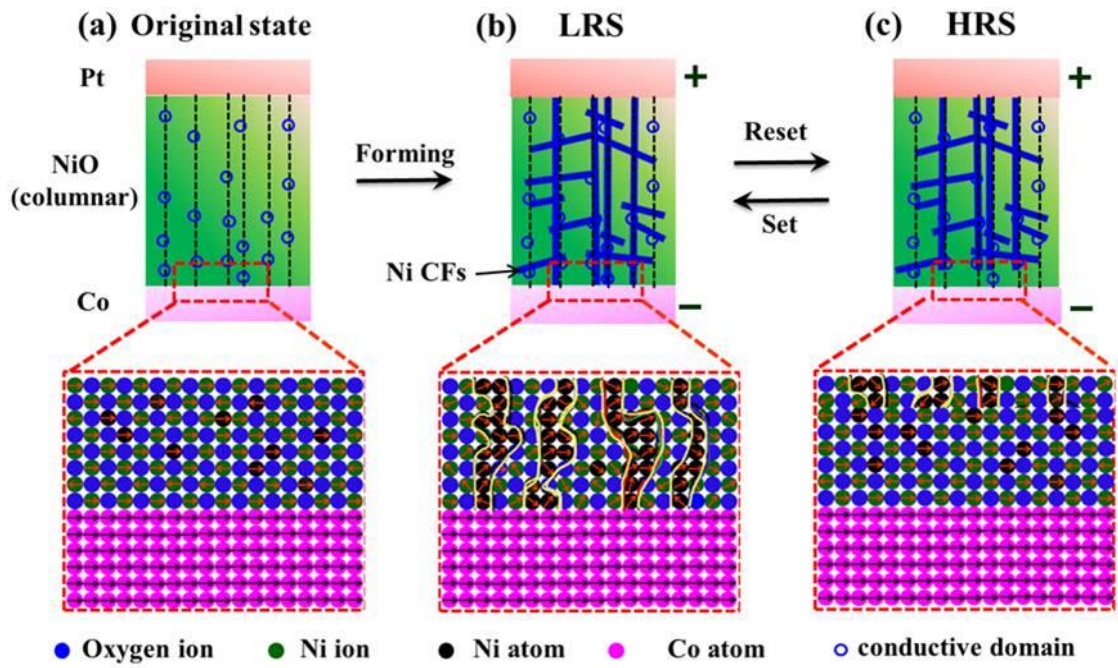

Fig. 5 Wei LJ et al